\documentclass[aps,prl,preprint,groupedaddress]{revtex4}

\usepackage{rotating}
\usepackage{amsmath}
\usepackage{graphicx} 

\begin{document}
\title{Black Hole Menagerie, Charged/Dyonic BHs and Radiation from Interacting Dyonic BH Pairs}









\author{Patrick Das Gupta}

\email[]{pdasgupta@physics.du.ac.in}

\author{Mohd. Sirtaz}

\email[]{sirtaz@physics.du.ac.in}



\affiliation{Department of Physics and Astrophysics, University of Delhi, Delhi - 110 007 (India)}








\begin{abstract}

Although black holes (BHs) as well as gravitational waves (GWs) were predicted from general relativity  more than hundred years back, it is only in recent times that substantial observational evidence is mounting up concerning their actual existence, thanks to LIGO/VIRGO detectors, Event Horizon telescope and tracking of stellar motion near the Galactic Centre. In this article, after discussing basic physics concerning classical BHs, we provide a pedagogical exposition of supermassive black holes, their relevance to  the understanding of active galactic nuclei, formation of  primordial black holes as well as  quantum aspects and challenges associated with BHs.

Thereafter, we describe charged BHs,  Penrose process for energy extraction from Kerr BHs and Wald's proposal concerning a Kerr BH slowly becoming a Kerr-Newman BH in the presence of a uniform magnetic field. In the context of BHs bearing magnetic charge, we discuss both magnetic monopoles as well as dyons, and their emergence from various models like string theory, GUTs and electroweak theories, etc. 

In the later portions, we concentrate on our  recent research work pertaining to the non-relativistic dynamics of dyon-dyon interaction that includes  mutual gravitational attraction. From the derived classical equations of motion, we obtain not only the well known Schwinger-Zwanziger quantization condition for dyons using Saha's argument based on quantized angular momentum of electromagnetic field but also a scalar virial theorem for an astrophysical system consisting of point particles, some of which carry both electric and magnetic charges. 

In the final sections, we obtain expressions for the generated electromagnetic wave  as well as gravitational wave amplitudes, and the corresponding luminosities due to dyon-dyon interactions. Lastly, we discuss the results after computing these quantities using a range of values for the mass, electric and magnetic charges, etc. of the dyonic BHs.

(A somewhat modified version of this article that includes several figures had been published in "A Guide to Black Holes" ed. Kenath Arun, Nova Science Publishers, 2022, ISBN: 979-8-88697-163-7)

\end{abstract}











%
%
%

\maketitle

\section{I. Introduction}

It is well established that special relativity not only placed  time and space  almost on the same footing,  it also  brought energy and momentum  together in a single framework, entailing  mass and energy to be identical, modulo a factor of $c^2$.  Einstein's general relativity (GR), a relativistic theory of gravitational interactions, that  rests firmly on the  pillars of  equivalence principle and special relativity, had gone  one step ahead and demonstrated that gravity is simply a manifestation of   space-time geometry. 
Einstein's equivalence principle  states that in a sufficiently `small'  freely falling  frame of reference, gravity disappears for a sufficiently `small' duration in this inertial frame, irrespective of the detailed nature of the gravitating source that is causing the free fall. The `smallness'  of such a local inertial frame, of course, depends on how strong, time varying  and non-uniform the local gravity  (i.e. local space-time geometrical curvature)  is. The space-time geometry in GR is determined by the energy and momentum densities as well as their flux densities  corresponding to all matter, radiation and  any other form of gravitating sources present in the universe.

One of the signature predictions of GR  is the existence of black hole (BH) solutions. A classical BH solution is characterized by an event horzon (EH) that surrounds the space-time singularity and  acts as a one-way boundary for all null-like  and time-like geodesics. The EH is analogous to a trapdoor that leads to an unending abyss. By characterizing space-time singularity  to be a point where all geodesics get terminated  (i.e. become  incomplete), Penrose had proved a powerful singularity theorem \cite{Pen} that had made use of the focussing effect of geodesics that ensue from  the Raychaudhari equation due to matter that satisfy strong energy condition \cite{Ray}. This theorem paved the way for the acceptance of singularities resulting from any kind of gravitational collapse involving  standard matter.

A BH singularity is  hidden  by an EH,  a mathematical surface encompassing the singularity. Any object that crosses the  EH towards the singular point cannot retrace its footsteps back or come out. For an  observer outside of the EH, time stands still on this one-way boundary  and any radiation emitted outward from this surface suffers infinite redshift, as long as one ignores quantum physics (for an extremely lucid introduction to BHs, see \cite{Vishu}). The idea that BHs are not merely mathematical solutions of GR but are generically created as end products of  stellar evolution in the universe, emerged  essentially from the pioneering research studies of Subrahmanyan  Chandrasekhar (for  pedagogical expositions, see e.g. \cite{srini1}, \cite{srini2}, \cite{pdg1}). From theoretical studies, the present understanding is that stellar mass BHs result from the collapse of stellar cores heavier than $\sim 3 - 10 \ M_\odot $.

The simplest BH case ensues from an exact solution of the Einstein equation, corresponding to the exterior of  a spherically symmetric mass distribution of finite radius $r_0$, obtained by Karl Schwarzschild in 1915, that was  published  in 1916, not long before he passed away (\cite{Karl}).  For $r>r_0$, as shown by Schwarzschild, the line-element for this exact solution has the following form,
\begin{equation}
ds^2= [1 - f(r)] c^2 dt^2 - \frac{dr^2}{1 - f(r)} - d\theta^2 - \sin^2{\theta} \ d\phi^2
\end{equation}
with  $f(r)$  given by the expression,
\begin{equation}
f(r)=\frac{R_s}{r}\ \ ,
\end{equation}
where the  Schwarzschild $R_s$  radius is defined as,
\begin{equation}
 R_s \equiv \frac{2 G M}{c^2 } \cong  3\ \mbox{km}\ \bigg (\frac{M} {M_\odot} \bigg )  \ ,
\end{equation}
  $M$ being the total mass of the spherically symmetric matter distribution. 

The Schwarzschild BH solution, with all its strangeness,  emerges from the above expressions naturally,  when the spherically symmetric object of mass $M$ shrinks to  a radius $r_0 < R_s$.  In such a situation, as we shall argue in the next section,  the entire matter making up the object  collapses unhaultingly  from $r_0$ to  $r=0$.  Hence,  matter density as well as the space-time curvature become infinite at $r=0$,  making the point a singularity. The EH corresponds to the surface $r= R_s$ with its  size being characterized  by the Schwarzschild radius $R_s$. As can be seen readily,  from eqs.(1) and (2), the metric components $g_{tt}$ and $g_{rr}$ become  0 and $- \infty$, respectively, on the EH.

In the realm of classical physics, a BH is uniquely characterized by its mass, angular momentum and charge (see e.g. \cite{Chandra}, and the references therein.). It does not matter whether a BH is created from the gravitational collapse of neutrons, anti-neutrons, photons, neutrinos/anti-neutrinos or any other  electrically neutral particle, as  it will  remember in this case, only the total mass and the total angular momentum that went in to form the BH,  ignoring blissfully the baryon number or the lepton number of the initial parent object. This follows from the standard `no hair' theorem for classical BHs.

In other words, unlike the macroscopic objects we encounter in our  day today lives, classical BHs are completely specified by just three physical quantities: $M$, $\vec{L}$ and $Q$.  Furthermore, as far as GR is concerned, BHs do not have any non-spacetime or material  structure whatsoever.  Mass, angular momentum and charge appear only as parameters  in the space-time metric of the BHs,  that determines the EH and the singularity. This is precisely the reason why Chandrasekhar had stated in his tome on BHs \cite{Chandra}:

{\it {`The	black	holes	of	nature	 are	the	most	perfect	macroscopic	objects there 	are	in	the	
universe:	the	only	elements	in	their	construction	are	our	concepts	of	space	and	time.	
And	since	the	general	theory 	of	Relativity	provides	only	a	single	unique	family	of	
solutions	for	their	descriptions,	they	are	the	simplest	objects	as	well.'}}

However, by considering quantum states of gravitons  corresponding to  BH exterior regions,  it has been  argued cogently  in a very recent paper that BHs  do possess `quantum hairs',  notwithstanding the above classical view of BHs \cite{Calmet}.

In the next section, we begin with  basic physics related to BHs and highlight their role in explaining the origin of very high luminosities from active galactic nuclei like quasars and blazars. Supermassive black holes as well as sub-solar mass primordial BHs, along with the puzzles and challenges associated with them, have  been discussed. We also dwell briefly on the subject of `Black hole atoms', Hawking radiation and the information loss paradox, in this section.

Section III describes charged BHs and Wald's proposed mechanism on how a Kerr BH, embedded in a uniform magnetic field,  can get  electrically charged. Thereafter, the topic of  magnetic monopoles and dyons has been discussed. Plausible origins of these hypothetical particles in string theory, Yang-Mills theories, Grand Unified theories (GUTs)  as well as  electroweak theory have been highlighted.

Our recent research work on the non-relativistic dynamics of dyon-dyon interaction has been taken up in section IV.  Apart from the electromagnetic forces between the dyons, Newtonian gravity between them has also been incorporated. In this section, we provide a detailed mathematical derivation of  the relevant dynamical equations as well as the ensuing  constants of motion, for a pair of dyons.  Saha's method of charge quantization in the presence of a magnetic monopole has been extended to the case of two static dyons in order  to obtain the Schwinger-Zwanziger quantization condition. Furthermore, a scalar virial theorem for a large gravitating system of point particles that includes a collection of dyons has been proved, in this section.

Section V covers the physics of field propagation in relativistic theories, in general, as well as the topic of gravitational radiation in general relativity. Electromagnetic radiation and gravitational waves resulting from dyon-dyon interactions have been elaborately dealt with in this section. 

In section VI, quantitative estimates of the electromagnetic wave  and gravitational wave amplitudes, as well as  the corresponding luminosities have been computed by considering various values for the mass, electric  and magnetic charges of an interacting electrically charged BH and a magnetically charged BH. Interestingly enough, for a given set of parameters, the EM radiation from such a pair of BHs mimics that of a Fast Radio Burst. Finally, section VII provides concluding remarks concerning our study.

\section{II. Black Hole Diversity: Supermassive, Primordial and Quantum Effects}

\subsection{BHs: Physical Aspects}

 Resuming our discussion from  the previous section on  the Schwarzschild solution, associated with a spherically symmetric mass,   eqs.(1) and (2)   imply that if the radius $r_0$ of this sphere is less than $R_s$ then a test particle in its vicinity  cannot be at a constant radius $r$ with $r_0 \ < r \  <\ R_s$. This is because, for a constant $r$  (i.e. $dr=0$) beneath the EH,
\begin{equation}
ds^2=[1 - \frac{R_s}{r}] c^2 dt^2   - d\theta^2 - \sin^2{\theta} \ d\phi^2 < 0 \ \ \ ,
\end{equation}
entailing any `constant $r$' world-line with $r < R_s$ to be  space-like. This prohibits any physical particle to be at a constant $r$ below $R_s$. 

In other words, even the surface matter of this spherical object (and thereby, the entire object) cannot be at a constant $r_0$, and instead, must contract 
 incessantly. This gravitating ball  cannot expand as that would violate energy conservation (see the discussion in the following paragraphs). Therefore, from the Schwarzschild  solution, it is evident that a spherically symmetric object of initial radius less than $R_s$ necessarily shrinks inexorably to the singular  point $r=0$.

It is a standard result in GR that if a space-time geometry admits a Killing vector $\xi^\mu $ then the quantity $\xi_\mu p^\mu $ is a constant of motion for a test particle having a  4-momentum $p^\mu$. For the Schwarzschild space-time described by eqs.(1) and (2), it is obvious that  $\xi^\mu= \delta^\mu_0$ is a time-like Killing vector since the metric is invariant under $t \rightarrow t + $constant. Hence, for a test particle of rest mass $m$, initially at rest, at the radial coordinate $r$, 
\begin{equation}
E = \xi_\mu p^\mu= g_{\mu \nu} \xi^\nu  p^\mu=  g_{0 0 } p^0= \sqrt{ 1- R_s/r} \  m c^2
\end{equation}
is its  conserved energy. From the above expression, at distances   $r \gg R_s$,  one finds that the conserved energy, $E \approx (1- R_s/2r)\   m c^2$ $ = mc^2  - GMm/r $, takes the expected form,  of being the sum of the test particle's rest energy  and its gravitational potential energy.

According to eq.(5), farther away the particle is located, greater is its energy. Thus, a particle at rest initially at $r_i$ can only fall inwards with $\dot {r}$ increasing with time, and not move upwards in the absence of  an external force acting on it.  Hence, a pressureless ball of dust-like matter can only contract. It is also interesting to note from the above equation that one may extract almost the entire rest energy $m c^2 $ of a test  particle by lowering it ever so slowly from $r_i \gg R_s$  down to the event horizon of a Schwarzchild BH since at $r \approx R_s$, $E \approx 0 $. 

 Therefore, very large luminosities and rapid time variability exhibited by the  high frequency EM radiation emitted by the  active galactic nuclei (AGNs) (e.g. blazars and quasars) can  very effectively be explained by invoking infall of matter from far away to regions close to the EH surface  of a supermassive black hole (SMBH). In fact, causality argument itself implies that  fluctuations in the accretion processes taking place close to the event horizon of the SMBH can give rise to variability time scales $\Delta t \sim \kappa \ R_s/c$, where  $\kappa $ is a dimensionless parameter $\gtrsim 1$.  For instance, typically  observed $\Delta t \sim $1 hour variability time scales in a blazar  would mean $R_s \sim 3.6 \times 10^{14}/\kappa \ $ cm, implying a SMBH of mass $\sim 10^ 8 M_\odot $.

 Similarly, one may estimate from eq.(5) the  maximum luminosity possible from matter forming an accretion disc,  and trickling down from an outer radius $\gg R_s$ slowly onto a Schwarzschild  BH due to viscous dissipation at a rate $\dot {m}$,
\begin{equation}
 L_{max} \cong \dot {m} c^2 \bigg [1- \sqrt {1 - \frac {R_s} {r_0}}\  \bigg ] \approx 0.18 \ \dot {m} c^2 
\end{equation}
 where  radius $r_0= 3 R_s $ is the innermost stable circular orbit  in the disc  (for very brief  introductions  to BH accretion,  see e.g. \cite{pdg1}, \cite{pdg2}). From eq.(6), one draws the conclusion that BH accretion can result in higher efficiency of  conversion of matter into radiation than nuclear fission or fusion. 

\subsection{Supermassive Black Holes}

 So far, most galaxies that have been studied closely reveal presence of  SMBHs with mass $\sim 10^6-10^7\ M_\odot$  at their centres. Very heavy SMBHs, on the other hand, are associated with power house of AGNs. This is because,  from eq.(6), high luminosity requires large values of accretion rate $\dot{m}$. But higher the luminosity, greater is the radiation pressure on the infalling matter  prohibiting the latter to be accreted at a rate more than the Eddington rate. Hence, in general, the AGN power is limited by the Eddington luminosity which  is proportional to the mass of the feeding BH, since larger the BH mass greater is its  inward gravitational force on the matter to counteract the outward force due to  radiation pressure.
Therefore, it is natural to expect a correlation between AGN luminosity and the mass of the underlying SMBH. 

The heaviest SMBH detected, as of now,  lies in the quasar TON 618, weighing  $\sim 66 \times 10^9\ M_\odot $. The light from this quasar that we see was emitted $\sim 10.4 \times  10^9\  $ yrs ago, when the  universe was only $\sim 3\times  10^9\  $ yrs old. So, how are such SMBHs created in the universe so early in latter's history? This is one of the biggest conundrums of cosmology. Several models have been proposed to address this puzzle (for a review,  see e.g. \cite{volon1}).

 Interestingly enough, if dark matter (DM) candidates instead of being  somewhat massive particles with rest energy more than 100 GeV, are actually  ultralight axion-like pseudoscalars with rest energy $m_{DM} c^2 \lesssim$ $ 10^{-20}$ eV, then a fraction of galactic mass scale DM halo can undergo  Bose-Einstein (BE) condensation. The quantum evolution of such BE condensates,  with  self-gravity included, can naturally lead to  creation of ultramassive BHs of mass $M_{SMBH}$,  satisfying a condition, 
\begin{equation}
m_{DM} \ M_{SMBH}  \ \gtrsim \ m^2_{Pl} \ \Rightarrow  \  \bigg (\frac {m_{DM} c^2}{10^{-20} \  \mbox{eV} } \bigg )  \bigg (\frac{M_{SMBH}}{10^{10} \ M_\odot} \bigg )\ \gtrsim  1 \ \  ,
\end{equation}
when the universe is barely billion years old \cite{pdg3, pdg4, pereira2021}. In eq.(7),  $m_{Pl} \equiv \sqrt{\hbar c/G} \approx 10^{-5}$ gm is the Planck mass. 

Origin of ultraheavy SMBHs when the universe is relatively young is enigmatic, as far as our theoretical understanding is concerned. On the other hand,  creation of sub-solar mass BHs (or primordial BHs)  in the radiation-dominated era of our universe had been theoretically predicted about five decades ago (see \cite{Hawking, Carr}, and also \cite{Zeldo} for a related idea).

\subsection{Primordial Black Holes}

 The argument delineated by Hawking and Carr as to why primordial BHs (PBHs) are expected to be formed in the early universe is remarkably simple \cite{Hawking, Carr}.  
At any epoch, the causally connected regions in an expanding universe is limited by  the Hubble radius, $R_H (t) \sim c/H(t) $, where the Hubble parameter is  given by $H(t)= \dot{a}/a$, $a(t)$ being the scale factor corresponding to a homogenous and isotropic cosmic model and $\dot{a} \equiv da/dt $ (for an elementary exposition, see e.g. \cite{pdg5}). In the radiation-dominated era, the scale factor goes as $a(t) \propto t^{1/2} $  so that $H(t) \sim 1/t$  and $R_H(t)  \sim c\ t$, while the highly relativistic particles that are in thermodynamic equilibrium have a temperature $T(t) \propto 1/a(t) \propto t^{- 1/2} $. 

Therefore, the energy density of the dominant components of the cosmic content is given by,
\begin{equation}
\rho(t) c^2 \sim a_B T^4 \sim \frac {3 c^2} {8 \pi G t^2}
\end{equation}
The mass enclosed within a Hubble radius then follows from eq.(8),
\begin{equation}
M_H(t) \sim \frac{4 \pi R^3_H} {3} \rho(t) \sim  \frac{4 \pi} {3}  (ct)^3 \rho(t)  \sim \frac {c^3 t} {2G} = 2 \times 10^3 \bigg ( \frac {t} {10^{-35}\ s} \bigg ) \ \ \mbox{gm} 
\end{equation}
From eq.(9), the Schwarzschild radius associated with $M_H(t) $ is,
 \begin{equation}
R^H_s (t) \equiv \frac { 2G M_H(t) } {c^2} \sim  \frac{2G} {c^2} \frac {c^3 t} {2G} = c t \sim R_H(t)\ !
\end{equation}
The scale $R_H$, characterizing the size of a causally connected region in an expanding universe, while $R_s$ providing the minimum size of a gravitating object below which it necessarily collapses into a BH, are two independent scales. So, it is rather surprising that they turn out to be comparable, as seen from eq.(10).   In other words, if at time $t$,  due to statistical fluctuations in the thermal energy density  of the relativistic components within a Hubble radius, the mass becomes  slightly greater than $ M_H(t) $ of eq.(9), the entire contents within the Hubble radius would collapse to form a PBH.

Of course, there are many other  theoretical models  proposed in the literature  concerning creation of   PBHs in the early universe e.g.  collision of bubble walls due to  first order phase transitions, collapse of  trapped false vacuum pockets, gravitational collapse of  inflaton/scalar fields/dark matter/dark energy, etc.  (see e.g. \cite{Carr1, Khlopov, pdg6, pdg7, pbh3}, and the references therein.).

Based on the observed extragalactic gamma-ray background as well as constraints arising from the Big Bang nucleosynthesis and the observed spectrum of the cosmic microwave background, stringent  restrictions on the mass fraction  as a function of the  PBH mass ensue (for a recent review, see \cite{Carr1}).
Here is an irony:    Observed super heavy SMBHs are fairly ubiquitous, although their origin so early in the cosmic history is still shrouded in mystery.  PBHs, on the other hand,  have so far evaded detection inspite of the well understood scenarios that would entail their formation.

\subsection{Black Holes and Quantum Theory}  

 Due to the predicted Hawking temperature of BHs \cite{Hawking1},   
\begin{equation}
T_H (M) = \frac { \hbar c} {4 \pi k_B R_s} = \frac {m^2_{Pl}  c^2} {8 \pi k_B M} \sim \frac {10^{26}}{M\ (\mbox{in  gm})}\ \ {}^\circ K \ ,
\end{equation}
associated with  the semi-classical Hawking  radiation resulting from the quantum effects of matter fields near the EH of a classical BH  of mass $M$,  PBHs  of mass $10^{14} - 10^{15}$ gm created more than billion years ago would be evaporating today, emitting gamma-ray photons copiously.  Moreover, PBHs created before the electroweak phase transition may also resolve the issue of baryogenesis in the early universe  provided there are adequate CP-violations in the decay of GUT bosons (see e.g. \cite{pdg6, pdg7}, and the references therein.). 

Now, through a gravitational capture in the early universe, a PBH of mass $M$ can certainly form a quantum mechanical bound state with an electron or a proton of mass $m$, with hydrogen-atom-like energy levels, 
\begin{equation}
E_n \approx  - \frac {G^2 M^2 m^3} { 2 \hbar^2  n^2}  = -  \frac {m c^2}{2 n^2} \bigg ( \frac {m c^2}{8 \pi k_B T_H(M)} \bigg )^2 \ ,
\end{equation}
 that follow from the energy eigenstates analysis and  eq.(11). From eq.(12),  it is easy to show that such a `Black hole atom' is unstable due to the ionizing effects  of its own Hawking radiation \cite{pdg2}.

Non-stop  Hawking evaporation of a BH until the latter's mass approaches Planck mass $m_{Pl} $, when quantum gravity effects are expected to take over,  leads to the well known BH information loss paradox \cite{Hawking1, Hawking2, Page}. This is essentially due to the fact that  a pure quantum state $|\psi^{BH}_0>$  describing a BH to begin with, becomes a mixed state that represents the thermal quantum state of the emitted particles due to  Hawking evaporation. Quantum theory's unitary evolution of state-vectors strictly forbids a pure state evolving into a mixed state.

 An elegant resolution of this paradox has been developed  by Samir Mathur and his group based on the string theory paradigm by demonstrating that for no microstate describing a BH in string theory, a conventional EH exists  and that, the usual BH space-time ends just before the EH.  Rather the microstates describing a stringy BH give rise  to a `fuzzy' horizon, and the  Hawking emission from such a `fuzzball' corresponds to a pure state,  without  violating unitarity,  not unlike the radiation from a thermally hot entity \cite{Samir, Samir1, Samir2, Samir3, Samir4}. 

On the other hand,  in 2013, the `firewall' proponents  Almheiri, Marolf, Polchinski and Sully had argued that  if the net Hawking radiation from a BH with a traditional EH  is indeed 
 in a pure state  then either the information carried by the emission is not from regions close to the EH  or  the EH itself manifests physically in the form of a  catastrophic discontinuity for a freely falling observer  \cite{Firewall}. 

 By providing a critical analysis of both the `fuzzball' as well as the `firewall' proposals, a comprehensive review of the  subject of Hawking radiation   has been put forward by Suvrat Raju  recently  that  includes his own insights pertaining to the information loss puzzle \cite{Suvrat}. According to Raju, careful calculations point to  the existence of a principle of redundancy entailing retention of the full copy of the BH interior information even outside the EH, and thereby, resolving the paradox. 

In other words, the jury is still out concerning the BH information loss paradox.

\section{III. Charged Black Holes, Monopoles and Dyons}

\subsection{Horizon, Ergosphere, Penrose Process and Wald's Mechanism}

The exact solution corresponding to the exterior metric of a spherically symmetric but electrically charged mass distribution, with total mass $M$ and total charge $Q$, were derived by Reissner in 1916 and, independently, by  Nordstr$\ddot{o}$m in 1918. Although the space-time  line-element has the same form as eq.(1), but because of the contribution of the electromagnetic field to the full energy-momentum tensor, the function  $f(r)$ is different from the Schwarzschild case, and is given by,
\begin{equation}
f(r)=\frac{R_s}{r}  - \frac{G Q^2} {c^4 r^2}
\end{equation}
so that, by setting $\frac{R_s}{r}  - \frac{G Q^2} {c^4 r^2}=0$, one obtains the EH for a Reissner-Nordstr$\ddot{o}$m BH to be at a radius,
\begin{equation}
R^{EH}_{RN}= \frac{GM}{c^2} + \sqrt{\bigg ( \frac{GM}{c^2}\bigg )^2  - \frac {G Q^2}{c^4}} 
\end{equation}
Since gravitation (or, equivalently space-time geometry) alone does not distinguish between $+Q$ and $-Q$ charge, it is not surprising that eqs.(13) and (14) have only quadratic dependence on the electric charge, and not a linear one. 

 From eq.(14), it is obvious that the EH size of a Reissner-Nordstr$\ddot{o}$m BH with  $Q \neq 0$ is always less than that of a Schwarschild BH of same mass, and that  it is necessary to have the condition  $G M^2 \geq Q^2 $  met,   for the solution to be valid (from a Newtonian point of view, the latter condition means that the net potential energy of the BH due to both gravity and Coulombic force must  be negative  for its stability). 

The most general classical BH, with mass $M$, angular momentum $J$ and electric charge $Q$, corresponds to the Kerr-Newman solution (see e.g. \cite{Chandra}), with its EH size being given by,
\begin{equation}
R^{EH}_{KN}= \frac{GM}{c^2} + \sqrt{\bigg ( \frac{GM}{c^2}\bigg )^2  - \bigg ( \frac{J}{M c}\bigg )^2 - \frac {G Q^2}{c^4}}   
\end{equation}
and its Hawking temperature by,
\begin{equation}
T^{KN} _H = \frac { \hbar c} {4 \pi k_B R^{EH}_{KN}} \frac {(R^{EH}_{KN})^2 -  (J/M c)^2 - \frac {G Q^2}{c^4}} { (R^{EH}_{KN})^2 + (J/M c)^2} \ ,
\end{equation}
which reduce to eqs.(3) and (11), respectively,  when $J=0=Q$. One recovers the corresponding Kerr BH expressions by setting $Q=0$ in eqs.(15) and (16).

As in the case of a Kerr BH, a Kerr-Newman BH too possesses  an ergosphere \cite{Ruffini} where test particles can have negative energy (as measured by inertial observers far away).  Suppose  a massive test particle  with an initial positive energy  $E$ enters the ergosphere, and then splits up into two particles A and B, with energy $E_A > 0$ and $E_B < 0$, respectively. Since Kerr BHs are stationary BHs,  admitting a time-like Killing vector $\xi^\mu = \delta^\mu_0 $, a particle's energy is conserved i.e.  $E= E_A + E_B$. As $E_B< 0$, it follows that $E_A > E$. Hence,  if the particle A emerges out of the ergosphere, one has mined energy out of the rotating BH at the expense of latter's rotational energy. This is the well known  `Penrose process' of  energy extraction \cite{Pen1, Pen2, Sanjeev}.

One may wonder how to prevent an electrically  charged BH  (even after creating it somehow) getting neutralized by attracting oppositely charged particles from an ambient plasma, since the former has an associated long range electric field. Interestingly enough, in 1974, Wald had provided a mechanism by which  a Kerr BH, embedded in  an ambient plasma as well as  a uniform magnetic field,  would eventually turn into a Kerr-Newman BH \cite{Wald}. 

Using the Killing vector equations ingeniously,  Wald had obtained an exact solution representing a Kerr BH of mass $M$ and spin angular momentum $\vec{J}$  along with  a magnetic field  surrounding it which,   at very large distances,  assumed a constant value  $\vec{B}_0$ (either parallel or anti-parallel to $\vec{J}$). Furthermore,  he had shown that  ambient charged particles of a given sign would be drawn preferentially  into the  BH, where  the sign of the charge depended on the sign of $(\vec{B}_0.\vec{J} )$. 


The charge $Q$ of such a Kerr-Newman BH would satisfy the constraint \cite{Wald},
\begin{equation}
\frac{Q} {(G M/c^2)^2} \leq 2  B_0  \ \ \ .
\end{equation}
Now, astrophysical BHs are all likely to be Kerr BHs, as majority of (progenitor) stars exhibit rotation.  Also, massive stars tend to be found in dense, gaseous environments.  In other words,   since magnetic field associated with cosmic plasma is ubiquitous,  there is a strong possibility of finding electrically charged BHs  in  the dense regions of stellar clusters, if one goes  by  Wald's solution. 

\subsection{Magnetic Monopoles and Dyons}

Electrically charged BH solutions can also be generalized to those with magnetic  and other gauge charges \cite{MagneticBH1, MagneticBH2, Sivaram, AstromagneticBHs}.  Although no magnetic monopole (i.e. a particle with a magnetic charge) has been seen in the nature so far,  the plausible existence of magnetic monopoles  not only enlarges the symmetry in electrodynamics, it also leads to a charge quantization condition \cite{Dirac, Saha}.  An electromagnetic duality transformation, a rotation in an internal space that mixes up both electric and magnetic fields as well as the electric and magnetic   charges in the same way,  is an exact symmetry of the extended Maxwell equations. 

 Furthermore, with quantum theory juxtaposed, the singular nature of the vector potential $\vec{A}(\vec{r},t)$ associated with a monopole with magnetic charge $g$ and the requirement that the wavefunction  describing a test particle with  electric charge $q$ be single valued, lead to the Dirac quantization, $q g=n \hbar c /2  $, thereby making the `magnetic  fine structure constant' $\alpha_{mag} \equiv g^2/\hbar c  > 34.25 $   \cite{Dirac, Preskill}.

 The concept of a dyon (a  hypothetical particle carrying both electric as well as magnetic charge,  $(q,g)$) was put forward by Schwinger in a paper that emphasized, among other interesting ideas,  on the U(1) global electromagnetic duality symmetry as well as on the ensuing large    $\alpha_{mag} $ \cite{Schwinger0}. Attempts to gauge the U(1)  electromagnetic duality symmetry have also been made in the past \cite{Pradhan, pdg8, pdg9}.

The main motivation of Schwinger's 1969 paper  was to provide a model  of hadronic matter being composed of dyons bearing fractional electric charges. This idea had an uncanny similarity with Saha's speculative paper of  1936 surmising that  neutrons, because of their non-vanishing magnetic dipole moment, are made of monopole-anti monopole pairs \cite{Saha}. In this paper, Saha also reproduced Dirac's seminal charge quantization rule  by invoking  quantization of the angular momentum of the electromagnetic field (see \cite{Saha, Wilson, Saha1, Goldhaber, Schwinger2, Wu}).  

Of course, in the string theory paradigm, interesting  mathematical solutions, representing dyons and dyonic BHs with  characteristic spectrum, have already been  obtained (see e.g. \cite{StringDyons, StringDyons1, StringDyons2, Ashoke, Ashoke1, DilatonicDyonBH}).  Similarly, magnetic monopole as well as dyon solutions, including dyonic BHs,   emerge naturally from non-Abelian gauge theories, with monopole mass as high as $\sim 10^{16}$ GeV, as in the case of GUT models \cite{Hooft, Polya, Nambu, Julia-Zee, cho1, cho2, DyonBHsYangMills, DyonBHsYangMills1, DyonBHsYangMills2}. 

What is perhaps more interesting is that because U(1) subgroup of SU(2)  in the Salam-Weinberg model of electroweak unification can have  non-trivial homotopy, the standard model entails presence of topologically stable magnetic monopoles weighing about 4 to 7 TeV (\cite{cho1}, \cite{cho2}, \cite{Ellis}). The estimated mass range makes the electroweak monopole quite appealing, as its existence can be corroborated  in the near future by  the MoEDAL detector at LHC (\cite{Acharya1}, \cite{Acharya2}).

Moreover, the electroweak phase transition that occurred when the universe was $\sim 10^{-11}$ s old and was very hot, corresponding to a temperature  $k_B T \sim  100$ GeV, would have led to production of magnetic monopoles with a net density of $\Omega_{0,\ mono}  \approx  6.5 \times 10^{-11}$ in the current epoch \cite{cho3}. 
However, for monopoles with mass less than about $10^{17}$ GeV there are strong observational constraints (\cite{Burdin}, \cite{Patrizii}).  So far, all searches for  monopoles and dyons have been  illusive,  yielding  null results.
 



\section{IV. Dynamics of Dyon-dyon Interaction}

\subsection{Equations of Motion and the Trajectories}

As an interesting theoretical problem, we begin  in this section, a study of dyon-dyon interaction. We consider  two dyons D1 and D2  in  an inertial frame S,  having electromagnetic charges $(q_1, g_1) $ and $(q_2, g_2) $, masses $m_1$, $m_2$ and instantaneous position vectors $\vec{r}_1 (t)$, $\vec{r}_2 (t)$, respectively. Apart from the electromagnetic interactions between them, their  mutual gravitational attractions have also been included in the present analysis. 


  

At an instant $t$ and at the position vector $\vec{r}$ of the frame S, the electric and magnetic fields sourced by   D1 as well as D2 in motion, are given by,
\begin{multline}
\vec{E} (\vec{r},t)= \vec{E}_1 (\vec{r},t)+ \vec{E}_2 (\vec{r},t) = \bigg [ q_1 \frac{\vec{r} - \vec{r}_1(t)}{\vert \vec{r} - \vec{r}_1(t) \vert^3} -  g_1 \frac{\dot{\vec{r}}_1 \times (\vec{r} - \vec{r}_1(t))}{c\vert \vec{r} - \vec{r}_1(t) \vert^3} \bigg ]  +  \bigg [ q_2 \frac{\vec{r} - \vec{r}_2(t)}{\vert \vec{r} - \vec{r}_2(t) \vert^3}  \\-  g_2 \frac{\dot{\vec{r}}_2\times (\vec{r} - \vec{r}_2(t))}{c\vert \vec{r} - \vec{r}_2(t) \vert^3} \bigg ]
\end{multline}
and,
\begin{multline}
\vec{B} (\vec{r},t)= \vec{B}_1 (\vec{r},t)+ \vec{B}_2 (\vec{r},t) = \bigg [g_1 \frac{\vec{r} - \vec{r}_1(t)}{\vert \vec{r} - \vec{r}_1(t) \vert^3}  +q_1 \frac{\dot{\vec{r}}_1 \times (\vec{r} - \vec{r}_1(t))}{c\vert \vec{r} - \vec{r}_1(t) \vert^3}\bigg ] +  \bigg [ g_2 \frac{\vec{r} - \vec{r}_2(t)}{\vert \vec{r} - \vec{r}_2(t) \vert^3}\\ +q_2 \frac{\dot{\vec{r}}_2\times (\vec{r} - \vec{r}_2(t))}{c\vert \vec{r} - \vec{r}_2(t) \vert^3} \bigg ] \ \ ,
\end{multline}
respectively.  In the above equations, $\vec{E}_1 $ ($\vec{E}_2$) is the electric field due to D1 (D2) while $\vec{B}_1 $ ($\vec{B}_2$) is the magnetic field generated by  D1 (D2).
(We have adopted the standard notations  $\dot{f} \equiv df/dt $ and $\ddot{f} \equiv d^2f/dt^2 $ throughout.)

It is fairly straightforward  to write down the equations of motion that govern the  classical dynamics of a pair of dyons (e.g. \cite{Zwanziger2}). What is new in this article is that we have extended the analysis  by incorporating the mutual gravitational attraction between the pair,
\begin{equation}
m_1 \ddot{\vec{r}}_1 = q_1 [\vec{E}_2 (\vec{r}_1,t) + \frac{\dot{\vec{r}}_1}{c} \times \vec{B}_2 (\vec{r}_1,t) ] +  g_1 [ \vec{B}_2 (\vec{r}_1,t) - \frac{\dot{\vec{r}}_1}{c} \times \vec{E}_2 (\vec{r}_1,t) ] - G m_1 m_2 \frac{\vec{r}_1 - \vec{r}_2}{\vert \vec{r}_1 - \vec{r}_2 \vert^3}
\end{equation}

\begin{equation}
 =   \frac{(- G m_1 m_2 + q_1 q_2 + g_1 g_2)(\vec{r}_1 - \vec{r}_2)}{\vert \vec{r}_1 - \vec{r}_2\vert^3} + \frac {(q_1 g_2 - q_2 g_1) (\dot{\vec{r}}_1 - \dot{\vec{r}}_2) \times (\vec{r}_1 - \vec{r}_2)}{ c \vert \vec{r}_1 - \vec{r}_2\vert^3 }  \ 
\end{equation}

and,
\begin{equation}
m_2 \ddot{\vec{r}}_2= q_2 [\vec{E}_1 (\vec{r}_2,t) + \frac{\dot{\vec{r}}_2}{c} \times \vec{B}_1 (\vec{r}_2,t) ] +  g_2 [ \vec{B}_1 (\vec{r}_2,t) - \frac{\dot{\vec{r}}_2}{c} \times \vec{E}_1 (\vec{r}_2,t) ] +  G m_1 m_2 \frac{\vec{r}_1 - \vec{r}_2}{\vert \vec{r}_1 - \vec{r}_2 \vert^3}
\end{equation}

\begin{equation}
 =   \frac{(- G m_1 m_2 + q_1 q_2 + g_1 g_2)(\vec{r}_2 - \vec{r}_1)}{\vert \vec{r}_1 - \vec{r}_2\vert^3} + \frac {(q_2 g_1 - q_1 g_2) (\dot{\vec{r}}_2 - \dot{\vec{r}}_1) \times (\vec{r}_2 - \vec{r}_1)}{ c \vert \vec{r}_1 - \vec{r}_2\vert^3 }  \ .
\end{equation}
In the present study,   D1 and D2 are assumed to be moving with non-relativistic speeds in S, so that $v_i \equiv \vert \frac {d \vec{r}_i} {dt} \vert \ll c,\ i=1,2$. Therefore, terms of the order  $\mathcal{O}(v^2_i/c^2)$  that appear in the dynamical equations have  been neglected.  For instance, the Lorentz force on D2 due to a  magnetic field $\vec{B}_{12} = \frac {q_1  \dot{\vec{r}}_1 \times (\vec{r}_2 - \vec{r}_1)} {c\vert \vec{r}_1 - \vec{r}_2\vert^3} $  generated by the motion of D1  at $\vec{r}_2$, etc. have been omitted since such terms are of $\mathcal{O} (v^2/c^2)$ order. One may refer to the study by Das and Majumdar for their analysis of charge-monopole scattering at Planckian energies \cite{DasMajumdar},  and \cite{Schwinger3} for non-relativistic dyon-dyon quantum scattering.

The separation between D1 and D2 represented by the relative vector $\vec r$, the position vector  of the centre of mass (CM)  $\vec R $ and the reduced mass are defined by, 
\begin{equation}
\vec{r}(t) = \vec{r}_2 (t) - \vec{r}_1 (t)\ , \  \  \vec{R}(t) = \frac {m_1\vec{r}_1 (t)+  m_2 \vec{r}_2(t)}{m_1+ m_2} \ ,  \ \ \mbox{and}\ \ \mu= \frac {m_1 m_2} {m_1 + m_2}   \ \ \  ,
\end{equation} 
respectively, so that eqs.(21) and (23) can be expressed in terms of  $\mu $, $\vec r$ and  $\vec R $ as follows,
\begin{equation}
\ddot{\vec{R}}=\vec{0} \Rightarrow \dot{\vec{R}}= \vec{V}_{CM}=\mbox{constant vector}\ ,
\end{equation}
and,
\begin{equation}
\mu\frac{d^2\vec{r}}{dt^2}=-\frac{1}{r^3}(b\vec{r}+a\vec{l}\:), 
\end{equation}
where,
$a \equiv (q_2g_1-q_1g_2)/(\mu c),\ b \equiv Gm_1m_2-(q_1q_2+g_1g_2)$ and $\vec{l}=\mu\vec{r}\times\dot{\vec{r}}$ is the angular momentum of the pair in the CM-frame.

It is easy to derive  three constants of motion ensuing  from eq.(26),
\begin{eqnarray}
E=\frac{1}{2}\;\mu\;\dot{r}\,^{2}+\frac{l^2}{2\mu r^2}-\frac{b}{r},\\
\vec{L}=\vec{l}-\frac{a\mu\vec{r}}{r}\\ \mbox{and  } l^2= L^2 -\mu^2 a^2  \ ,
\end{eqnarray}
where $E$ and $\vec{l}$ are the  energy and the angular momentum, respectively, of the reduced system in the CM frame. In the CM-frame, $\vec{V}_{CM}$ is, of course,  identically zero.

Making use of eq.(28), eq.(26) may be re-expressed as,
\begin{equation}
\frac{d^2\vec{r}}{dt^2}=-\frac{1}{\mu\ r^3}\bigg [\bigg (b\ +\ \frac{\mu a^2}{r} \bigg) \vec{r} \ +\ a \vec{L} \bigg ]  \ \ .
\end{equation}
Since $\vec{L}$ remains constant, we may take the z-axis to be along this constant vector.  Then, adopting spherical polar coordinates and making use of eqs.(28) and (29), we have with us,
\begin{equation}
\vec{r}.\vec{L}= r L \cos{\theta_0}= - a \mu \ r \Rightarrow \cos{\theta_0} = - \frac{a \mu}{L} \mbox{ and } \sin{\theta_0} = \pm \frac{l}{L} \ ,
\end{equation}
implying that the  vector $\vec{r}(t) $, representing the separation between the two dyons,  always lies on a cone of half-opening angle $\theta_0$ with its apex at the origin.

 From pure kinematics, i.e. by expanding $\vec{l}\times \vec{r}$, one can also derive an useful expression,
\begin{equation}
\mu \dot{\vec{r}} =\frac{1} {r} \bigg (\mu \ \dot{r} \ \vec{r} - \frac{\vec{r}}{r} \times  \vec{l} \bigg )
=\frac{1} {r} \bigg (\mu \ \dot{r} \ \vec{r} - \frac{\vec{r}}{r} \times  \vec{L} \bigg )
\end{equation}
so that, with $\vec{L}= L \hat{e}_z$, we get an expected result,
\begin{equation}
\dot{z} = \frac {\dot{r}} {r} z \ \ \Rightarrow\ \  z(t)= r(t) \cos{\theta_0}
\end{equation}
Moreover, with this choice of the z-axis, we obtain from eq.(30),
\begin{equation}
\ddot{x} = - \frac {1} {\mu\ r^3} \bigg [b\ +\ \frac{\mu a^2}{r}\bigg ] x \ \  ,
\end{equation}
and,
\begin{equation}
\ddot{y} = - \frac {1} {\mu\ r^3} \bigg [b\ +\ \frac{\mu a^2}{r}\bigg ] y \ \ ,
\end{equation}
so that, $x = r \sin{\theta_0} \cos{\phi}$ and $y = r \sin{\theta_0} \sin{\phi}$ leads to the instantaneous angular speed,
\begin{equation}
\omega(t)= \dot{\phi}= \pm\ \frac{L} {\mu\ r^2} \ \ .
\end{equation}
Above considerations allow one to immediately obtain circular orbit solutions ($\dot{r}=0$ and $r(t)=r_0$ = constant) of eq.(26).  In this case, it is trivial to see from eq.(33) that $z_0= - a \mu r_0/L $ constitutes the orbital  plane. The radius of the circular orbit is given by $\rho_0 = \sqrt{r^2_0 - z^2_0}= l r_0/L$. Employing eqs.(27) and (36), one obtains for the  orbit,
\begin{eqnarray}
r_0= \begin{cases}\frac{l^2}{2 \mu b} \ \Rightarrow b> 0,&E=0\\ \frac{l^2}{ \mu b}\Big [1 \ \pm \ \sqrt{1\ +\ 2l^2 E/\mu b^2}\Big ]^{-1},&E \neq 0
\end{cases}
\end{eqnarray}
and a constant angular speed,
\begin{equation}
\omega = \pm\ \frac{L} {\mu\ r^2_0}
\end{equation}

\subsection{Charge Quantization Condition for Dyons}

Saha's argument for charge quantization can be  extended trivially to the case of two static dyons  in order to arrive at the Schwinger-Zwanziger quantization condition \cite{Schwinger1, Zwanziger2, Zwanziger1}. Referring to Fig.1 as well as eqs.(18) and (19), if one considers the dyons to be at rest (i.e.  $\dot{\vec{r}}_1=0=\dot{\vec{r}}_2 $), the net electric and magnetic fields at $\vec{r}_P$ due to D1 and D2 are given by,
\begin{equation}
\vec {E} (\vec{r}_P, t)= q_1 \frac {(\vec{r}_P - \vec{r}_1 (t))} {\vert \vec{r}_P - \vec{r}_1 (t) \vert^3} + q_2 \frac {(\vec{r}_P - \vec{r}_2 (t))} {\vert \vec{r}_P - \vec{r}_2 (t) \vert^3}
\end{equation}
and,
\begin{equation}
\vec {B} (\vec{r}_P, t)= g_1 \frac {(\vec{r}_P - \vec{r}_1 (t))} {\vert \vec{r}_P - \vec{r}_1 (t) \vert^3} + g_2 \frac {(\vec{r}_P - \vec{r}_2 (t))} {\vert \vec{r}_P - \vec{r}_2 (t) \vert^3}
\end{equation}
so that the Poynting vector is proportional to,
\begin{equation}
\vec {E} (\vec{r}_P, t) \times \vec {B} (\vec{r}_P, t) = (q_1 g_2 - q_2 g_1) \frac{(\vec{r}_P - \vec{r}_1 (t)) \times (\vec{r}_P  - \vec{r} _2 (t))}{\vert \vec{r}_P - \vec{r}_1(t) \vert^3 \vert \vec{r}_P - \vec{r}_2 (t) \vert^3}  
\end{equation}
The EM field angular momentum due to two static dyons then is simply,
\begin{equation}
\vec {l}_{EM} (t)= \frac {1} {4 \pi c} \int {\vec{r}_P \times (\vec {E} (\vec{r}_P, t)\times \vec {B} (\vec{r}_P, t)) d^3 r_P}
\end{equation}
After integrating over $\vec{r}_P$, one obtains,
\begin{equation}
 \vec {l}_{EM} (t) = - \frac {q_1 g_2 - q_2 g_1}{c} \ \ \frac {\vec{r}_2 (t) - \vec{r}_1 (t)} {\vert \vec{r}_2 (t) - \vec{r}_1 (t) \vert} =  - \frac {q_1 g_2 - q_2 g_1}{c} \ \ \frac {\vec{r}(t)} {r(t)} 
\end{equation}
(The  above integration can be performed trivially by assuming D1 to be at the origin of the coordinate system, without any loss of generality, so that $\vec{r}_1 =\vec{0}$.)

Applying now Saha's argument that angular momentum of the electromagnetic field must be quantized as $\vert \vec {l}_{EM} \vert = n \hbar/2 $, one reaches the desired  result,
\begin{equation}
 \frac {q_1 g_2 - q_2 g_1} {c} = n \hbar /2  \  ,
\end{equation}
namely,  the well known Schwinger-Zwanziger quantization condition.

\subsection{Scalar Virial Theorem for a Self-Gravitating System\\ Containing Dyons}

Suppose one is interested in  astrophysical BHs acquiring both electric and magnetic charges by  accreting dyons statistically from the environment. For such a possibility, the stellar mass  BHs  must belong to a large virialized system since the  time scale  on which the BHs can become charged, in a stochastic manner,  is expected to be of the order of billion years. Hence,  it is instructive to take a closer look at the virial equilibrium condition for a self-gravitating system consisting  of baryonic matter, dark matter, stellar mass BHs as well as dyons. For simplicity, we adopt a point particle description for  each of the constituents of such a  astrophysical cloud. 

  We begin our study with a large cosmic nebula consisting of $N$ randomly moving particles/dyons characterized by their position vectors $\vec{r}_i(t) $, masses $m_i$ and electromagnetic charges $(q_i, g_i)$, $i=1,2,...,N$.  For the ones that are electrically neutral, $q_j=0=g_j$.  Eqs.(21) and (23) can be extended in a straightforward manner  to express the force experienced by the $j^{th}$ particle/dyon due to  rest of the particles,
\begin{multline}
\vec{F}_j= m_i \ddot{\vec{r}}_j =\sum\limits_{{{k = 1,k}} \ne {{j}}}^N {\frac{1}{{{{\left| {{{\vec r }_{j - }}{{\vec r }_k}} \right|}^3}}}\left[ {( - G{m_j}{m_k} + {q_j}{q_k} + {g_j}{g_k})({{\vec r }_j} - {{\vec r }_k})} \right.}\\
\left. + {\frac{{{q_j}{g_k} - {q_k}{g_j}}}{c}({{\mathop {\vec r }\limits^ \cdot  }_j} - {{\mathop {\vec r }\limits^ \cdot  }_k}) \times ({{\vec r }_j} - {{\vec r }_k})} \right]
\end{multline}
Taking the dot product of eq.(45) with $\dot{\vec{r}}_j$ and then summing over all the particles, we get,

{\footnotesize{\begin{align}
&\sum^N_{j=1} \sum^N _{k=1, k\neq j} { \dot{\vec{r}}_j. \vec{F}_j}=  \frac{d}{dt}\sum^N_{j=1} \sum^N _{k=1, k\neq j} {\frac{1} {2} m_i \dot{\vec{r}}_j.\dot{\vec{r}}_j} \equiv \frac {dT}{dt}=
\\
&= \sum^N_{j=1} \sum^N _{k=1, k\neq j}  \nonumber\\
& {\frac{ \dot{\vec{r}}_j  \ .} {\vert \vec{r}_j - \vec{r}_k\vert^3}  \bigg [(- G m_j m_k + q_j q_k + g_j g_k)(\vec{r}_j - \vec{r}_k) +  \frac {q_j g_k     - q_k g_j}{c} (\dot{\vec{r}}_j - \dot{\vec{r}}_k) \times (\vec{r}_j - \vec{r}_k) \bigg ]}
\end{align}}}
Considering half of the RHS of eq.(47) with  indices $j$ and $k$ interchanged as well as using the anti-symmetry of $q_j g_k - q_k g_j$ in  $j$ and $k$, we may express the rate of change of the kinetic energy of the system as,
\begin{multline}
\frac {dT}{dt}= \frac {1}{2} \sum^N_{j=1} \sum^N _{k=1, k\neq j}  {\frac{ (\dot{\vec{r}}_j -  \dot{\vec{r}}_k)\ .} {\vert \vec{r}_j - \vec{r}_k\vert^3}}  
\bigg [(- G m_j m_k + q_j q_k + g_j g_k)(\vec{r}_j - \vec{r}_k) +  \frac {q_j g_k - q_k g_j}{c} (\dot{\vec{r}}_j \\
-{ \dot{\vec{r}}_k) \times (\vec{r}_j - \vec{r}_k) \bigg ]}
\end{multline} 

The dot product of $\dot{\vec{r}}_j -  \dot{\vec{r}}_k$ with the second term of the above equation  vanishes identically, reflecting the fact that   work done by magnetic (electric) fields on electric (magnetic) charges is zero. 

For any vector $\vec{a}$, we have,
 $$ \frac{d}{dt} \bigg (\frac{1}{|\vec{a}|} \bigg )= - \frac {\vec{a}.\dot{\vec{a}}}{|\vec{a}|^3} \ , $$ and hence, eq.(48) can be expressed as,
\begin{equation}
\frac {dT}{dt} =   - \frac {d}{dt}\Big (\frac {1}{2} \sum^N_{j=1} \sum^N _{k=1, k\neq j}  {\frac{- G m_j m_k + q_j q_k + g_j g_k} {\vert \vec{r}_j - \vec{r}_k\vert}}\Big ) = - \frac {dV}{dt} 
\end{equation}

$$\Rightarrow \  E \equiv T + V \ \mbox{is the conserved energy, where, }$$
\begin{equation}
 V \equiv \frac {1}{2} \sum^N_{j=1} \sum^N _{k=1, k\neq j}  {\frac{- G m_j m_k + q_j q_k + g_j g_k} {\vert \vec{r}_j - \vec{r}_k\vert}}
\end{equation}
is the potential energy of the cloud.

Now, the centre of mass position vector is simply, $\vec{R} \equiv \frac {\sum^N_i  {m_i \vec{r}_i}} {\sum {m_i}} = \frac {\sum^N_i {m_i \vec{r}_i}} {M}$, where $M=\sum^N_i {m_i}$ is the total mass of the cosmic cloud. Using eq.(45) one can  easily demonstrate that,
\begin{equation}
\ddot{\vec{R}} = \frac {1} {M}\sum {m_j \ddot{\vec{r}}_j}\vspace{-10pt}
\end{equation}
\begin{multline}
=\frac {1} {M} \sum^N_{j=1} \sum^N _{k=1, k\neq j}  {\frac {1} {\vert \vec{r}_j - \vec{r}_k \vert ^3}}\bigg [(- G m_j m_k + q_j q_k + g_j g_k) (\vec{r}_j - \vec{r}_k) +  \frac {q_j g_k - q_k g_j}{c} (\dot{\vec{r}}_j \\
{- \dot{\vec{r}}_k) \times (\vec{r}_j - \vec{r}_k) \bigg ]}=0
\end{multline}
because of the anti-symmetry of the force term that is being summed over with respect to  the indices $j$ and $k$. Therefore, the kinetic energy associated with centre of mass motion, $T_{CM} = \frac{1} {2} M \dot{R}^2$ is a constant of motion so that, from eqs.(49) and (50), we have,\vspace{-10pt}

{
\footnotesize{\begin{equation}
E - T_{CM}= T - T_{CM} + V = \sum^N_{i=1}  {\frac{1} {2} m_i \dot{\vec{r}}_i . \dot{\vec{r}}_i - \frac{1} {2 }  M \dot{\vec{R}}.\dot{\vec{R}}} + V = \frac{1} {2 M} \sum^N_{i=1}\sum^N_{j=1}{m_i m_j ( \dot{\vec{r}}_i . \dot{\vec{r}}_i - \dot{\vec{r}}_i . \dot{\vec{r}}_j}) + V 
\end{equation}}}
\begin{equation}
= \frac{1} {2 M}  \sum^N_{i=1}\sum^N_{j=1}{m_i m_j \dot{\vec{r}}_i .(\dot{\vec{r}}_i - \dot{\vec{r}}_j)} + V =  \frac{1} {4 M}  \sum^N_{i=1}\sum^N_{j=1}{m_i m_j (\dot{\vec{r}}_i - \dot{\vec{r}}_j).(\dot{\vec{r}}_i - \dot{\vec{r}}_j)} + V
\end{equation}
to be also a constant of motion. In particular, if $T - T_{CM} + V < 0$, the self-gravitating cloud  would be a bound system. Of course, this necessitates $V < 0$. 

One may also obtain a scalar virial theorem for such a  massive object made of dyon-particle-BH  mixture. Given the moment of inertia, $I=\sum^N_{i=1} {m_i \vec{r}_i. \vec{r}_i}$, equations of motion entail \cite {Chandra1},
\begin{equation}
\frac{1}{2} \frac{d^2 I} {dt^2} = 2 T + \sum^N_{j=1} \vec{r}_j . \vec{F}_j  \ \ .
\end{equation}
In our case, making use of eq.(45), the second term of the RHS of eq.(55) reduces to,
\begin{multline}
\sum^N_{j=1} \vec{r}_j . \vec{F}_j= \sum^N_{j=1} \sum^N_{k=1, k \neq j} {\frac{\vec{r}_j } {\vert \vec{r}_j - \vec{r}_k\vert^3}} . \bigg [(- G m_j m_k + q_j q_k + g_j g_k) (\vec{r}_j - \vec{r}_k) +  \frac {q_j g_k - q_k g_j}{c}  (\dot{\vec{r}}_j\\ 
{- \dot{\vec{r}}_k) \times (\vec{r}_j - \vec{r}_k) \bigg ]} 
\end{multline}\vspace*{-44pt}

\begin{multline}
= \frac {1}{2} \sum^N_{j=1} \sum^N_{k=1, k \neq j} {\frac{(\vec{r}_j - \vec{r}_k).} {\vert \vec{r}_j - \vec{r}_k\vert^3}} \bigg [(- G m_j m_k + q_j q_k + g_j g_k) (\vec{r}_j - \vec{r}_k) +  \frac {q_j g_k - q_k g_j}{c} (\dot{\vec{r}}_j\\ 
{- \dot{\vec{r}}_k) \times (\vec{r}_j - \vec{r}_k) \bigg ]} = V   \ \ ,
\end{multline}
where $V$ is given by eq.(50).

If we define,
\begin{equation}
I_0 \equiv I - M \vec{R}.\vec{R}
\end{equation}
then, using eqs.(52), (53), (55) and (57), it can be easily shown that the scalar virial theorem takes the form,
\begin{equation}
\frac{1} {2} \ddot{I}_0 = \frac{1} {2} \ddot{I}- M \dot{\vec{R}}.\dot{\vec{R}} = 2 T + V - M \dot{\vec{R}}.\dot{\vec{R}}= 2 (T-T_{CM}) + V 
\end{equation}

In the continuum limit, the potential energy,
\begin{equation}
 \sum^N_{i=1} \sum^N _{j=1, i\neq j}  {\frac{- G m_i m_j + q_i q_j + g_i g_j} {\vert \vec{r}_i - \vec{r}_j\vert}} 
\end{equation} 
becomes,
 \begin{equation}
\rightarrow V= \int {d^3 r \ \int {d^3 r'  \  \frac{- G \rho (\vec{r}, t) \rho (\vec{r'}, t)   + \rho_{el} (\vec{r}, t) \rho_{el} (\vec{r'}, t)  + \rho_{mag} (\vec{r}, t) \rho_{mag} (\vec{r'}, t)} {\vert \vec{r} - \vec{r}' \vert}}}
\end{equation}
where $\rho$, $\rho_{el}$ and $\rho_{mag}$ are the mass, electric and magnetic charge densities, respectively.

For a near spherical, uniformly dense cloud of radius $\sim R$, mass $M$, net  electric charge $Q_e$ and net magnetic charge $Q_m$, the above expression leads to,
\begin{equation}
V \sim \frac {- G M^2 + Q^2_e + Q^2_m} {R}
\end{equation}
so that, using eqs.(59) and (62),  the conditions for the cosmic ball to be in a bound  state as well as  in virial equilibrium (i.e.  $\ddot{I}_0=0$) take the form,
\begin{equation}
Q^2_e + Q^2_m < G M^2 \mbox{  and  }  <T>  - \  T_{CM} \ \cong  \ - \ \frac{1} {2} <V> \ \ \ .
\end{equation}

As long as the conditions given by eq.(63) are met, a cosmic cloud containing a large number of dyons, among other entities like stars and BHs,  can certainly be in a state of virial equilibrium. BHs (initially neutral), lying in such  astrophysical clouds,  can absorb  dyons statistically and  become electromagnetically charged  gradually.  For instance, if a BH randomly accretes $N_{\mbox{dyon}}$ number of dyons, its  electromagnetic charge will eventually be proportional to $\sqrt{N_{\mbox{dyon}}}$  (see e.g. \cite{MagneticBH2}, and the references therein). 

\section{V. Radiation from Interacting Dyons: Electromagnetic and Gravitational Waves}

In any theory of fields that is consistent with special relativity (SR), the field perturbations would propagate  as  waves with speeds $v \leq c$. This is because,  all physical perturbations must  bear    energy, since only those entities that carry energy can  be detected or measured, for the simple reason that  exchange of energy between an entity and a sensing device  is necessary for the former's  detection. Even in quantum physics this is true since a quantum system and a measurement apparatus  can influence each other only through a suitable interaction Hamiltonian.  Now, an energy carrying perturbation has an associated mass via $E=mc^2$ relation. Hence, because of the result $ m = m_0/\sqrt{1- v^2/c^2}$  of SR,  $m_0$ being the rest mass,   the propagation speed of any measurable perturbation being $ \leq c$  is a stringent requirement in any physical theory.

Both electrodynamics as well as GR are consistent with SR. Thus, it ensues that electromagnetic (EM) waves and gravitational radiation (i.e.  undulations
in space-time geometry)  travel in vacuum with speed $c$.   Existence of gravitational waves (GWs)  predicted by GR  was indirectly 
 corroborated by the meticulous  observations  of the  binary pulsar system PSR 1913 + 16 by Taylor and Hulse, who demonstrated that the orbital period of  PSR 1913 + 16  decreased in accordance with the loss of orbital energy due to the emission of GWs (\cite{Damour}, and the references therein). The first ever direct detection of GWs from a binary black hole system involving $36\ M_\odot$ and $29\ M_\odot$ BHs was achieved by LIGO in 2015 \cite{Abbott}.

Emission of EM waves and GWs from  pure electrically charged BH binary was studied  by Leibling and Palenzuela in 2016 \cite{Leibling and Palenzuela}.   The primary objective of this section is to consider such radiation from  a pair of interacting dyonic BHs, moving non-relativistically. Since we are interested in studying orbiting heavy dyon pairs of astrophysical significance,  the precise nature of their orbital trajectories needs to be studied.  From eq.(27), the rate at which the distance between the pair changes is given by,
\begin{equation}
\dot{r}= \pm \sqrt{\frac{2}{\mu} \Big (E + \frac{b} {r} - \frac{l^2}{2 \mu r^2}\Big)}
\end{equation}
The above equation can  easily be integrated to yield, 
\begin{equation}
t=K\pm\frac{1}{ E }\sqrt{\frac{\mu}{2}}\Big[\sqrt{E r^2 +br-\frac{l^2}{2\mu}}-\frac{b}{2}I_1\Big] 
\end{equation}
where, 
\begin{eqnarray}
I_1 \equiv \begin{cases}\frac{1}{\sqrt{E}}\ln\Big(2\sqrt{E(Er^2+br-\frac{l^2}{2\mu})}+2Er+b\Big),&E>0\\ \frac{1}{\sqrt{E}}\sinh^{-1}\Big(\frac{2Er+b}{\sqrt{\Delta}}\Big),&E>0,\Delta>0\\- \frac{1}{\sqrt{-E}}\sin^{-1}\Big(\frac{2Er+b}{\sqrt{-\Delta}}\Big),&E<0,\Delta<0
\\ \frac{1}{\sqrt{E}}\ln\Big(2Er+b\Big),&E>0,\Delta=0
\end{cases}
\end{eqnarray}
and $-\Delta  \equiv b^2+\frac{2l^2E}{\mu}$, with  $K$ being a  constant of integration.

Expressing eq.(28) as,
\begin{equation}
\mu \vec{r} \times \dot{\vec{r}} = \vec{L} + \frac{a\mu\vec{r}}{r} \ \ \ ,
\end{equation}
one can obtain the following equivalent result by taking a cross product of the above equation with $\vec{r}$,
\begin{equation}
\mu \dot{\vec{r}} =\frac{1} {r} \bigg (\mu \dot{r} \vec{r} - \frac{\vec{r}}{r} \times  \vec{L} \bigg )
\end{equation}
so that one may use eqs.(64) and the constancy of  $ \vec{L} $ in the above to solve for $\vec{r} (t)$.


Choosing the  $z$-axis to be in the direction of $\vec{L}$ (hereafter, we will assume that $\vec{L} = L \hat{z}$ throughout the paper) and making use of eq.(65), we can integrate eq.(68) to obtain,
{\footnotesize{\begin{eqnarray}
x(t)=r (t) \sqrt{1-\alpha^2}\,\cos(\phi (t) + K_1),y(t)=r (t) \sqrt{1-\alpha^2}\,\sin(\phi (t) + K_1),z (t)=\alpha\,r (t)\\
\alpha \equiv -a\mu/L,\;\phi(t) \equiv \pm\frac{L}{\sqrt{2\mu}}\begin{cases}\frac{\sqrt{2\mu}}{l}\sin^{-1}\Big(\frac{br-l^2/\mu}{r\sqrt{-\Delta}}\Big),&\Delta<0\\ \frac{\sqrt{2\mu}}{l}\tan^{-1}\Big(\frac{\sqrt{\mu/2}(br-l^2/\mu)}{l\sqrt{Er^2+br-\frac{l^2}{2\mu}}}\Big),\\-\frac{2\sqrt{br+Er^2}}{br},&l=0

\end{cases}
\end{eqnarray}}}

\noindent with $K_1$ being a constant of integration and $r(t)$ obtained from eq.(65). The instantaneous angular frequency is given by,
\begin{equation}
\omega (t) = \frac{d\phi} {dt}= \frac{L}{\mu r^2 (t)}
\end{equation}
Because of the non-central nature of the net force mediated between the pair, the motion is confined to the surface of a cone (eqs.(31), (69) and (70))  instead of a plane.  If the pair is bound, i.e. $E<0$, one can easily demonstrate from eqs.(65) and (66) that there are two turning points corresponding to $r_1 $ and $r_2$, representing the minimum and maximum separation between the pairs, respectively.

As long as the closest distance of approach between the dyonic BHs is much larger than a scale determined by their gravitational radii  i.e. $r_1 > 10 \ R_{BH} = 2 G m_2/c^2$,  $m_2$ being the larger of the two BH masses, their speeds would remain non-relativistic and, hence,  would be amenable to the above analysis.

In what follows, we take up the case of  EM waves and GWs emission by a pair of dyons, and obtain the formal expressions for amplitudes and powers that  make use of the above equations. Thereafter, by considering various parameter sets, we tabulate the estimated  quantities numerically from the expressions derived theoretically, in the following subsections.

\subsection{V (a). Electromagnetic Dipole Radiation}

Since from eqs.(65), (66), (69) and (70), the trajectories of D1 and D2 have been  determined analytically, it is quite straightforward to calculate the EM wave amplitude and the power resulting from the interacting pair. 

Given the electric and  magnetic dipole moments,
$\vec{d}=q_1\vec{r}_1+q_2\vec{r}_2$ and $\vec{m}=g_1\vec{r}_1+g_2\vec{r}_2$, respectively, one can show using eqs.(21) and (23) that,
\begin{equation}
\ddot{\vec{d}} = - \frac{m_1q_2 - m_2 q_1}{m_1 m_2 r^3} \bigg [b\vec {r} +  \bigg (\frac{\mu a^2 \vec{r}}{r} + a \vec{L} \bigg )\bigg ] 
\end{equation}
and,
\begin{equation}
\ddot{\vec{m}} = - \frac{m_1g_2 - m_2 g_1}{m_1 m_2 r^3} \bigg [b\vec {r} +  \bigg (\frac{\mu a^2 \vec{r}}{r} + a \vec{L} \bigg )\bigg ]  \ ,
\end{equation}
so that the retarded electric  and magnetic field amplitudes  at the point of observation $\vec{r}_{\mbox{obs}} $, very far away from the pair of dyons, are given by (\cite{Landau}),
\begin{equation}
\vec{E}(\vec{r}_{\mbox{obs}}, t)=\frac{1} {c^2 r_{\mbox{obs}}} [(\ddot{\vec{d}}\times \hat{n} )\times \hat{n} + \hat{n} \times \ddot{\vec{m}}]_{t - {r}_{\mbox{obs}}/c}
\end{equation} 
and 
\begin{equation}
\vec{B}(\vec{r}_{\mbox{obs}}, t)= \frac{1} {c^2 r_{\mbox{obs}}} [(\ddot{\vec{m}}\times \hat{n} )\times \hat{n} - \hat{n} \times \ddot{\vec{d}}]_{t - {r}_{\mbox{obs}}/c}   \  \  .
\end{equation}

Making use of eqs.(72) and (73) in  eqs.(74) and (75), one obtains for the electromagnetic wave amplitude the following expressions,
\begin{align}
\vec{E}(\vec{r}_{\mbox{obs}}, t)&=\frac{1} {r_{\mbox{obs}} c^2} \Big[ \Big (\frac{b+\frac{\mu a^2}{r}}{r^3} \Big ) \Big((Bx-Ay)\hat{y} - (Ax+By)\hat{x}\Big ) \Big]_{t - {r}_{\mbox{obs}}/c}  \\ 
\vec{B}(\vec{r}_{\mbox{obs}}, t)&= \hat{n} \times \vec{E}(\vec{r}_{\mbox{obs}}, t)
\end{align}
where $A=q_1/m_1-q_2/m_2,B=g_1/m_1-g_2/m_2$, and that the z-axis has been chosen along the conserved vector $\vec{L}$.

 The instantaneous power radiated in the form of electromagnetic waves is given by (\cite{Landau}),
\begin{equation}
P_{em}=\frac{2}{3c^3}(\ddot{\vec{d}}.\ddot{\vec{d}} + \ddot{\vec{m}}.\ddot{\vec{m}})
\end{equation}
which in our case, from eqs.(72) and (73) along with $\vec{L} = L \hat{z}$, reduces to,
\begin{equation}
P_{em}=\frac{2}{3c^3r^6}(A^2+B^2)(a^2l^2+b^2r^2)
\end{equation}

We will use the expressions  given by eqs.(76), (77) and (79) to estimate the EM fields and luminosities in  section VI  after we choose a set of parameters in order to  tabulate and discuss our results.







%
%
%
%
%
%

\subsection{V (b). Gravitational Radiation}


   
At very large distances from a source for which energy-momentum distribution varies in a  non-relativistic manner, the gravitational  perturbation  represented by $h_{\mu \nu} $  is  expected to be  weak and thus one may choose a quasi-Minkowskian coordinate system to express the metric tensor as,
\begin{equation}
 g_{\mu \nu} \approx \eta_{\mu \nu} + h_{\mu \nu} (\vec r, t) \  ,
\end{equation}
with the perturbation amplitude satisfying,
\begin{equation}
\vert h_{\mu \nu} (\vec r, t) \vert  \ll 1  \ .
\end{equation}
As general covariance is built into GR, out of the 10 components of $h_{\mu \nu} $ only 2 propagating modes are physical since 6 of them can be made to vanish by making suitable  general coordinate transformations \cite{Schutz, Maggiore}. It is these two propagating modes that correspond to a GW amplitude. 

The standard practice is to adopt a harmonic coordinate condition or  the transverse, traceless (TT) gauge in which the GW amplitude,  made up   of  a linear combination of only the space-space components  of $h_{\mu \nu} $ representing  two distinct polarizations   $h_+ $ and $h_\times $,  is  normal to the direction of propagation.  For instance,  if  a   GW is propagating in the z-direction,  it may have only $h^{TT}_{11}= - h^{TT}_{22} =  h_+ $ and $h^{TT}_{12}= h^{TT}_{21}= h_\times $ as the non-zero components. Clearly, the GW amplitude then is manifestly traceless and transverse to the z-direction.

The GW amplitude, in the TT-gauge,  as measured  by an observer at time t and at a distance  $ {r}_{\mbox{obs}} $  from the source, with ${r}_{\mbox{obs}}  \gg $  size of the source,  is given by,
\begin{equation}
h^{TT}_{ij} (t, \vec {r}_{\mbox{obs}}) = \frac{2 G} {c^4  r_{\mbox{obs}} }  \frac {d^2 Q^{TT}_{ij} (t -  r_{\mbox{obs}}/c)}{dt^2}  \  ,
\end{equation}
where $Q^{TT}_{ij}$ is obtained from the mass quadrupole moment of a source with mass density $\rho(t, \vec r)$,
\begin{equation}
Q^{ij} (t) \equiv \int {\rho(t, \vec r) x^i x^j d^3r} 
\end{equation}
by applying suitable projection operators on eq.(83) \cite{Schutz}.

The energy-momentum pseudo-tensor corresponding to GWs is given by \cite{Landau},  
$$T^{GW}_{\mu \nu}=\frac {c^4} {32\pi G} \bigg < h^{TT}_{jk, \mu} h^{TT \  jk}_{,\nu} \bigg > $$
where $h^{TT}_{jk}$ is the GW amplitude in the TT-gauge with  $<...>$ representing an  average  over many wavelengths (raising and lowering of indices of GW amplitude are done using the Minkowski metric tensor $\eta_{\mu \nu}$.).

 The basic hurdle in obtaining a `proper' energy-momentum tensor for the gravitational degrees of freedom is due  to the fact that the metric tensor $g_{\mu \nu}\rightarrow \eta_{\mu \nu}$ and the Christoffel symbol $\Gamma^\mu_{\alpha \beta}\rightarrow 0$   in a local inertial frame (i.e. in a freely falling frame),  S,  so that any second rank tensor built using $g_{\mu \nu}$ and $\Gamma^\mu_{\alpha \beta}$ will vanish at an  event point $\cal{E}$  in S. But this point $\cal{E}$   is arbitrary as one is free to choose a local inertial frame anywhere in the whole of the  space-time manifold. This implies that such a tensor is identically zero every where. Using first derivatives of $\Gamma^\mu_{\alpha \beta}$,   symmetric tensors like $R_{\mu \nu}$ or $G_{\mu \nu}$ can definitely be constructed. However, $R_{\mu \nu}$ and $G_{\mu \nu}$ vanish where there is no matter because of the Einstein equations, and thus,  cannot represent  energy and momentum flux of GWs traversing through vacuum. Hence, one falls back on the Landau-Lifshitz energy-momentum pseudo-tensor to study the energy and momentum associated with the gravitational degrees of freedom \cite{Landau}.

As $T^{GW}_{00}$ represents  the GW energy density, the GW energy flux is given by,
$$F_{gw}= c T^{GW}_{00}= \frac {c^5} {32\pi G} \bigg < h^{TT}_{jk, 0} h^{TT\ jk}_{,0} \bigg >=
\frac {c^3} {32\pi G} \bigg < \dot h^{TT}_{jk} \dot h^{TT\ jk} \bigg >  $$
where  $\dot h_{jk} \equiv \frac {\partial h_{jk}} {\partial t}$. By using  eq.(82) in the above  equation one obtains,
\begin{equation}
F_{gw}= \frac {1} {4 \pi r^2_{\mbox{obs}}} \bigg (\frac {G} {2 c^5}  \bigg < \dddot{{Q}^{TT}_{jk}} \dddot{Q}^{TT\ jk} \bigg > \bigg )\ .
\end{equation}
Now,  the  surface area of a sphere of radius $r$  is $\cong 4 \pi r^2 $ if  the background space-time is nearly flat.  So, when the emission of GWs from a source is nearly isotropic,  eqs.(80), (81)  and (84)  entail that the GW power be  given by, 
\begin{equation}
P_{gw}\cong 4 \pi r^2_{\mbox{obs}} \ F_{gw}=\frac {G} {2 c^5}  \bigg < {\dddot{Q}^{TT}_{jk}} \dddot{Q}^{TT\ jk} \bigg > \ \ .
\end{equation}
For the pair of dyonic BHs, the mass quadrupole moment $Q^{ij}$ is given by,
\begin{equation}
Q^{ij} = m_1 x^i_1 x^j_1  + m_2 x^i_2 x^j_2 \  ,
\end{equation}
so that when one uses the relative separation vector  $\vec r $ and the position vector $\vec R$ of the CM as given by eq.(24), the above equation takes the form,
\begin{equation}
Q^{ij} =  (m_1 + m_2)  X^i X^j  +  \mu x^i  x^j \  .
\end{equation}
Hence, the time derivative of the mass quadrupole moment is simply,
\begin{equation}
\dot{Q}^{ij} =  (m_1 + m_2)  [\dot{X}^i X^j  + X^i \dot{X}^j ] +  \mu [\dot{x}^i  x^j  + x^i  \dot{x}^j] \  .
\end{equation}
From eq.(25), we had earlier concluded that the velocity of the CM of the pair is a constant. Hence, if we choose an inertial frame S' that  is comoving with the CM, so that  $ \vec R = \vec 0$ and $\vec{V}_{CM} = \vec 0$,  the expressions for  $ Q^{ij} $ and  $\dot{Q}^{ij} $ (eqs.(87) and (88)) get  simplified considerably to,
\begin{equation}
Q^{ij} =  \mu x^i  x^j \  ,
\end{equation}
and,
\begin{equation}
\dot{Q}^{ij} =   \mu [\dot{x}^i  x^j  + x^i  \dot{x}^j] \  .
\end{equation}


In the CM-frame, therefore,  the higher time derivatives of the mass quadruple moment tensor, after making use of eq.(26),  are given by,
\begin{equation}
\ddot{Q}_{ij}=-\frac{1}{r^3}\Big[a(l_ix_j+l_jx_i)+2bx_ix_j\Big]+2\mu\dot{x}_i\dot{x}_j \ ,
\end{equation}
and,

{\footnotesize{
\begin{equation}
\dddot{Q}_{ij}=\frac{3\dot{r}}{r^4}\Big[a(l_ix_j+l_jx_i)+\frac{2}{3}x_{i}x_{j}\Big(3b+\frac{\mu a^2}{r}\Big)\Big]-\frac{1}{r^3}\Big(4b+\frac{\mu a^2}{r}\Big)\Big(\dot{x}_{i}x_j+\dot{x}_jx_i\Big)\\-\frac{3a}{r^3}\Big(l_i\dot{x}_j+l_j\dot{x}_i\Big) \ .
\end{equation}}}
Suppose we choose the z-axis of the frame S' to be along $\vec L$ (note that  $\vec L$ is a constant of motion (eq.(28)).  Then,  for the GW propagating along the z-axis the TT-gauge 
gravitational wave amplitudes corresponding to the two distinct polarizations are  given by,
\begin{equation}
h_+ (t, \vec{r}_{\mbox{obs}}) =\frac{1}{r_{\mbox{obs}}}\frac{G}{c^4}(\ddot{Q}_{xx}-\ddot{Q}_{yy}) (t - {r}_{\mbox{obs}}/c) \ ,
\end{equation}
\begin{equation}
h_{\times}(t, \vec{r}_{\mbox{obs}})=\frac{2G}{r_{\mbox{obs}}c^4}\ddot{Q}_{xy} (t - {r}_{\mbox{obs}}/c)\ ,
\end{equation}
where $\ddot{Q}_{xx},\,\ddot{Q}_{yy}$ and $\ddot{Q}_{xy}$ are to be calculated  using eq.(91) along with eqs.(65), (66), (69) and (70).

 The gravitational wave power ensuing from eq.(85) has the form \cite{Schutz, Maggiore, PeterMathews},
$$P_{gw}=\frac{G}{5c^5} \bigg (\sum^2_{i=1}{\sum^2_{j=1}{(\dddot{Q}_{ij} \dddot{Q}_{ij}-\frac{1}{3}\dddot{Q}_{ii} \dddot{Q}_{jj}) }} \bigg )\  ,$$
so that 
\begin{equation}
P_{gw}=\frac{G}{5c^5}\Big[\frac{b^2}{r^4}\Big(\frac{8}{3}\dot{r}^2+32\frac{l^2}{\mu^2r^2}\Big)+\frac{2a^2l^2}{r^6}\Big\{\frac{a^2}{r^2}+\frac{9l^2}{\mu^2r^2}+\frac{8b}{\mu r}\Big\}\Big] \ ,
\end{equation}
after making use of eq.(92).

When we set $(q_i, g_i)=0$, $i=1,2$, the expressions given by eqs.(91)-(95) reduce to those derived by Peter and Mathews \cite{PeterMathews}, in their seminal paper describing the generation of  gravitational radiation from a pair of bound uncharged mass points.  

In the next section, we compute the values of GW amplitudes as well as the corresponding power for various sets of parameters.

\section{VI. Results and Discussions}

If dyons that are predicted from several high energy physics theories, do exist in our universe, it is extremely crucial that measurable effects resulting from dyon-dyon interactions are studied quantitatively so that they can be verified in the near future using gravitational wave detectors, radio-telescopes and other astronomical techniques. Therefore, in this section, we consider a range of values for the parameters that determine the amplitudes and luminosities pertaining to the electromagnetic as well as gravitational radiation from a pair of bound dyons, D1 and D2. For simplicity, in our computations, we have assumed  D1 to carry  only magnetic charge and D2 to bear electric charge alone. In any case, because of the electromagnetic duality symmetry of the extended Maxwell equations, one can always choose a duality rotation angle such that a given dyon carries only an electric charge (or, for that matter, only a magnetic charge).

Since, in this study,  we have restricted ourselves to  non-relativistic dynamics for the pair D1 and D2, we have considered only those cases in which the closest distance of approach $r_1 > 18\ R_{BH}$, where $R_{BH} \equiv 2 G m_2/c^2$ with $m_2 > m_1$, and an initial relative speed $v=0.01\ c$ for calculating the value of the angular momentum $l$ in the CM-frame. As the BHs, in our case,  are charged, the corresponding EH radii are smaller than $R_{BH}$ and, hence, the condition $r_1 > 18\ R_{BH}$ ensures that the Newtonian approximations are valid through out  the orbital motion executed by the pair.

The results have been summarized in the tables. As expected, it is evident from the tables, that GW luminosities increase when one chooses larger values of masses while the electromagnetic charges decide the magnitude of EM wave power. From table I and table II, one notices a very interesting feature in that, when the BH masses are $\sim 10^{26}$ gm,  electric charge for D2 is $q_2 \sim 10^{-5}\  \sqrt{G}\ m_2 - 10^{-3}$ $\sqrt{G}\ m_2$ and magnetic charge for D1 is $g_1= \hbar c/2q_2$,  in consonant with Dirac quantization condition, the emanating EM power as well as the characteristic frequency are similar to what one observes for typical fast radio bursts (for a very short review on fast radio bursts (FRBs), see \cite{pdg10}). Of course, the standard explanations for FRBs  hinge on astrophysical neutron stars endowed with very high magnetic fields $|\vec B | \gtrsim 10^{15}$ Gauss, lying amidst magnetized plasma.

Because of two distinct kinds of forces acting between the bound pair (electromagnetic and gravitational), there appears in the dynamics two different types of periodic motion. One is the fast gyration characterized by the instantaneous angular frequency $\omega $, given by eqs.(36) and (71), arising due to the varying Lorentz force. The other is the non-planar periodic motion due to a combination of electrostatic and gravitational forces.  Fig.2 and Fig.7  display the time evolution of the separation vector $\vec r$ between D1 and D2 from when they were  maximally separated ($r_2$) to the closest distance of approach ($r_1$), corrresponding to half of the slow periodic motion. The rapid gyration determined by $\omega $ is clearly reflected in the figures. 

From choosing various values for the parameters, one finds that the wave amplitudes as well as the luminosities are very large when D1 and D2 approach $r_1$. This is expected since $\omega $ and  the velocities increase sharply  with decreasing separation $r$, entailing rapid changes in electromagnetic dipole moments as well as mass quadrupole moments. However, in the present analysis, we have not included the radiative reactions in our calculations. That will be the subject matter in an upcoming  research paper of ours.

\section{VII. Conclusions}

The subject of black holes (BHs) has come a long way since the times when they were considered  only as  exotic mathematical solutions ensuing from Einstein's general relativity.   After about  mid-1970s onwards, indirect evidence for their actual presence in the cosmos began pouring in  from the X-ray observations of galactic stellar mass BH candidates such as   Cygnus X-1 as well as from the multi-wavelength studies of numerous  active galactic nuclei like quasars, blazars and radio-galaxies that confirmed the involvement of these ever hungry,  supermassive, gravitational behemoths in their central engines. 

 `Near direct' evidence for BH existence started mounting since 2015 with the gravitational wave detectors LIGO/VIRGO becoming operational, leading to direct observations of gravitational radiation  from an ever increasing number of BH binary systems.   Of course,  the 2020 Physics Nobel Prize  itself was awarded to Roger Penrose for his pioneering contributions to the singularity theorems that entail BHs to form inevitably in gravitational collapse of  sufficiently massive systems as well as to Andrea Ghez and Reinhard Genzel for demonstrating  independently  that the gravity monster  lying at the center of the Milky Way,  which is responsible for the  telltale stellar trajectories in the vicinity of  Sagittarius A*, in all likelihood can only be a supermassive BH weighing about 4.5 million times the Sun's mass. Furthermore,   recent   radio imaging of the photon spheres by the Event Horizon Telescope provided quantitative mapping of the space-time geometry just outside the event horizon of the supermassive black holes (SMBHs) located at the centers of M87 and Milky Way galaxies, respectively.

From the perspective of theoretical physics,  diverse high energy physics models   such as  string theory, grand unified theories (GUTs), models describing  unification of electromagnetic and weak forces, etc., that are based on various kinds of fundamental quantum fields, have been leading to mathematical solutions describing  exotic BHs of microscopic size  that carry electromagnetic as well as other types of charges. So, in the current research milieu   that encompasses fundamental physics and astrophysics, there is no escape from BHs.  The objective of this chapter, therefore, has been to provide  an understanding of  the basic underlying physics pertaining to a zoo of BHs and physical phenomena associated with them.

In the earlier sections of the chapter, one saw why the event horizon that encloses a BH singularity acts as a one-way traffic surface, and  why  SMBHs act as  highly efficient engines in  converting  matter into radiation,  eventually  leading  to the bizarre observed behavior of quasars, blazars, etc. Kerr BHs, which are pure gravitational, space-time whirlpools,   and a proposed Penrose process for  extracting their rotational energy were also briefly covered. Similarly, challenges associated with SMBH formation, production of primordial black holes in the early universe as well as their quantum aspects like Hawking radiation and the ensuing information loss paradox, were dealt with in these sections.

Can  BHs be electrically charged? Can they be magnetically charged? Such questions along with  Wald's proposal whereby a spinning BH turns gradually  into an electrically charged  BH in the presence of a uniform magnetic field were discussed.   Since BHs may harbor  magnetic charge if magnetic monopoles exist, topics pertaining to magnetic monopoles,  dyons (hypothetical particles bearing both electric as well as magnetic charges)  and dyonic BHs that  emerge invariably as mathematical solutions in different high energy physics theories like GUTs,  electroweak unification, string theory, etc., were  also included in the deliberation. Schwinger-Zwanziger quantization condition for dyons, using M. N. Saha's argument based on quantized angular momentum of electromagnetic field,  as well as  a scalar virial theorem for a gravitating cosmic cloud  consisting of point particles, some of which could be  dyons,  were  derived in the later portions of the chapter.

How does one find out whether such dyons or dyonic BHs exist in the real world? One way is to look for characteristic radiation from a pair of interacting dyons. Hence, in the final sections of the chapter,  a recent  study of ours  on dyon-dyon interactions, with gravity switched on, was  considered and the expressions for the generated electromagnetic wave as well as gravitational wave amplitudes, and the corresponding radiated power, that follow from the theoretical analysis,  were obtained  in a rigorous manner. The relevant physical quantities were  computed using a range of values for the mass, electric and magnetic charges, etc. of the dyonic BHs in order that their characteristic signatures may be verified or constrained  later from future observations.

\section{Acknowledgements}
PDG thanks Professor Debades Bandyopadhyay and Professor Debasish Majumdar, Saha Institute of Nuclear Physics, Kolkata, who had  invited him to present the above research work of ours   in   the month of January, 2020, at the AAPCOS 2020 confererence. It is also a pleasure to thank Professor Ashoke Sen, Professor Samir Mathur, Professor Debasis Ghoshal and Professor Suvrat Raju for their valuable comments during several workshops and conferences. 

\newpage
   
\vskip 1 em
\noindent {\bf Tables: Radiative Power and Characteristic Frequency - Electromagnetic and Gravitational Waves}
\vskip 1  em
\renewcommand{\arraystretch}{0.5}
\begin{table}[h]\label{table-1}
\caption{Average electromagnetic wave power, gravitational radiation power and characteristic  frequency from a bound system, consisting of a magnetic BH and an electric BH    with charge to mass ratio:  $\frac {q_2} {\sqrt{G} m_2} = 1.0\times10^{-5} $ $ \mbox{ and magnetic charge:} \;g_1= \frac{\hbar c } {2 q_{2}} \ ; $ \;  $R_{bh} \equiv 2 Gm_2/c^2$.}
\centering
\resizebox{\textwidth}{!}{
\begin{tabular}{|c|c|c|c|c|c|c|c|c|}
\hline

\multicolumn{6}{|c|}{Parameters}&\multicolumn{3}{|c|}{Results}\\

\hline
$m_1$ & $m_2$&$r_1/R_{bh}$& $r_2/R_{bh}$&$E\;\mbox{(erg)}$&$l\; (g\;cm^2 s^{-1})$&$\omega \;(\mbox{rad s}^{-1} )$ & $P_{em}\; (\mbox{erg s}^{-1} )$ & $P_{gw}\; (\mbox{erg s}^{-1} ) $\\
\hline
$29.0 \ M_\odot $& $36.0 \ M_\odot$&$20.6$& $617.4$&$-4.0\times10^{52}$&$6.1\times10^{52}$&$40.0$&$ 2.5 \times 10^{40}$&$ 1.5 \times 10^{50}$\\
\hline
$5.0\ \times 10^{26}\mbox{g}$&$5.0\ \times 10^{26}\mbox{g}$&$18.6$&$557.4$&$ -3.9\times10^{44}$&$3.4\times10^{36}$&$7.0\times10^9$ &$ 5.7 \times 10^{40}$&$ 4.2 \times 10^{50}$\\
\hline
$2.9\ \times 10^{21}\mbox{g}$&$3.6\ \times 10^{22}\mbox{g}$&$34.4$&$1031.7$&$-1.2\times 10^{39}$&$2.6\times10^{27}$&$2.8\times10^{13}$ &$ 3.2 \times 10^{37}$&$ 6.8 \times 10^{46}$\\
\hline
$1.0\times10^{15}\mbox{g}$&$1.0\times10^{16}\mbox{g}$&$33.8$&$1013.4$&$-4.3 \times10^{32}$&$2.4 \times10^{14}$&$1.0\times10^{20}$ &$ 5.2 \times 10^{37}$&$ 1.2 \times 10^{47}$\\
\hline
\end{tabular}
}
\end{table}
\begin{table}[h]\label{table-2}
\caption{Average electromagnetic radiation power, gravitational wave   power and characteristic  frequency from a bound system consisting of a magnetic BH and an electric BH   with charge to mass ratio:  $\frac{q_2} {\sqrt{G} m_2} = 1.0\times10^{-3} \ \; \mbox{and magnetic  charge:} \;g_1= \frac{\hbar c } {2 q_2} \ ;$ \; \ $R_{bh} \equiv 2 Gm_2/c^2$.}
\centering
\resizebox{\textwidth}{!}{%

\begin{tabular}{|c|c|c|c|c|c|c|c|c|}
\hline

\multicolumn{6}{|c|}{Parameters}&\multicolumn{3}{|c|}{Results}\\

\hline
$m_1$ & $m_2$&$r_1/R_{bh}$& $r_2/R_{bh}$&$E\;\mbox{(erg)}$&$l\; (g\;\mbox{cm}^2 s^{-1})$&$\omega\; (\mbox{rad s}^{-1} )$ & $P_{em}\; (\mbox{erg s}^{-1} )$ & $P_{gw}\; (\mbox{erg s}^{-1} ) $\\
\hline
$5.0\times 10^{26}\mbox{g}$&$6.0\times 10^{26} \mbox{g}$&$20.3$&$608.1$&$-3.5\times10^{44}$&$4.4 \times10^{36}$&$5.0\times 10^9$&$2.8 \times 10^{44}$ & $ 1.7 \times 10^{50}$ \\
\hline
\end{tabular}}
\end{table}
\begin{table}[h]\label{table-3}
\caption{Average electromagnetic radiation power, gravitational wave   power and characteristic  frequency from a bound system consisting of a magnetic BH and an electric BH   with charge to mass ratio:  $\frac{q_2} {\sqrt{G} m_2} = 1.0\times10^{-3}\ \; \mbox{ and magnetic charge:} \;g_1= \frac{\hbar c } {2e}\ ,\; \mbox{where the }\; \mbox{electronic charge, } e=4.8\times10^{-10}\; \mbox{esu}$\ ;\; \ $R_{bh} \equiv 2 Gm_2/c^2$.}
\centering
\resizebox{\textwidth}{!}{%

\begin{tabular}{|c|c|c|c|c|c|c|c|c|}
\hline

\multicolumn{6}{|c|}{Parameters}&\multicolumn{3}{|c|}{Results}\\

\hline
$m_1$ & $m_2$&$r_1/R_{bh}$& $r_2/R_{bh}$&$E\;\mbox{(erg)}$&$l\; (g\;\mbox{cm}^2 s^{-1})$&$\omega\; (\mbox{rad s}^{-1} )$ & $P_{em}\; (\mbox{erg s}^{-1} )$ & $P_{gw}\; (\mbox{erg s}^{-1} ) $\\
\hline
$10^{16}\mbox{GeV}/c^2$&$10^{20}\mbox{g}$&$38.0$&$682.0$&$-1.1\times10^{6}$&$4.7 \times10^{-9}$&$5.0\times 10^{17}$&$1.0 \times 10^{6}$ & $ 9.1 \times 10^{-9}$ \\

\hline
\end{tabular}}
\end{table}
   \begin{itemize}
 \item {\bf Set 1:} Values of the different parameters used in our analysis:
   \begin{enumerate}
   \item $g_1=\sqrt{G}m_1 10^{6}, g_2=0, q_1=0, q_2=\sqrt{G}m_2 
   10^{-4}, m_1=29 M_\odot, m_2=36 M_\odot , b \equiv Gm_1m_2-(q_1q_2+g_1g_2) $ 
   \item $R_{bh}=2Gm_2/c^2, \ r_{\mbox{obs}} =410\ \mbox{Mpc} ,\ r_{2} =617.43\ R_{bh}    	,\  r_{1}=20.6\ R_{bh}$
   \item $\phi_0=\frac{\pi}{2},l= 600 \ \mu\,v\,R_{bh},  \; v=0.01\ c , \;E=\ - \ \frac{\mu b^2}{16l^2}$   
 \end{enumerate}
   \end{itemize}

\begin{itemize}
\item {\bf Set 2:} Values of the parameters considered in our study:
   \begin{enumerate}
   \item Electronic charge $e=4.8\times10^{-10}\ \mbox{e.s.u.}, g_1=\frac{hc}{4\pi e}, g_2=0, q_1=0, q_2=\sqrt{G}m_2 
   10^{-5}, m_1=10^{16}\ \mbox{GeV}/c^2, m_2=10^{20}\ \mbox{gm},  b \equiv Gm_1m_2-(q_1 q_2+g_1 g_2) $ 
   \item $R_{bh}=2Gm_2/c^2, r_{\mbox{obs}}=10^{13}\ \mbox{cm}, r_{2} =682 \ R_{bh}  ,  r_{1}=38 \ R_{bh}.$
   \item $\phi_0=\frac{\pi}{2} , l=600 \ \mu\,v\,R_{bh}, \; v=0.01\ c , \;E=\ - \  \frac{\mu b^2}{10l^2}.$
 \end{enumerate} 
   \end{itemize}


\end{document}